\documentclass{edp-conf}
\usepackage{graphicx}

\begin{document}

\TitreGlobal{SF2A 2004}


\title{Initial power spectrum shape}
\author{Douspis, M.}\address{LATT/OMP, 14 av. E. Belin, 31400 Toulouse}
%
\runningtitle{Initial power spectrum shape}
\setcounter{page}{1}
\index{Douspis, M.}

\maketitle

\begin{abstract} Mainly used to set constraints on cosmological parameters, 
CMB anisotropies can give information on primordial
universe. Unfortunately, initial conditions and late time cosmology
are degenerated. I focus on the effect of the shape of the initial
power spectrum of fluctuations on the cosmological parameter
estimation from CMB temperature anisotropies. Strong constraints
obtained using a priori knowledge of the initial spectrum disappear
when the latter is let free.
\end{abstract}

%

\section{Introduction}

 CMB anisotropies have been used widely to put constraints on
 cosmological parameters (CP hereafter) during the last ten
 years. Until recently, the inhomogeneous set of observations and
 systematic effects have lead to limited constraints on some
 parameters (or combinations of parameters). Most of these limitations
 seem to disappear with the new WMAP data which encompass a large sky
 coverage (full sky) and large scale range (from arc minute to 180
 degrees, or in Fourier modes, from $\ell=2$ to
 $\ell=800$). Nevertheless, some degeneracies are intrinsic to CMB
 physics and can not be broken even with a perfect experiment (when
 considering temperature anisotropies only). I focus in this work on
 the degeneracy between initial conditions and late time cosmology and
 in particular between the shape of the initial power spectrum and the
 matter content of the universe.

\section{Shape of the initial power spectrum}

\subsection{Power law spectrum}

 The shape of the initial power spectrum (IPS hereafter) is not known
 a priori and is often assumed to be a power law caracterised by an
 amplitude $A$ and an index $n$ ($P \propto A k^n$). The case $n=1$ is
 referred as Harrison--Zel'dovitch spectrum.  If such assumption is
 made (in addition to the flatness, trivial topology of the universe
 and unicity of adiabatic initial modes), one can retrieve with good
 accuracy most of the cosmological parameters (see Spergel et
 al. 2003 and Fig.~1 (right)).

\subsection{Spectrum by intervalle}

 One can also find that a unique amplitude for all scale is a bit
 simple and parametrize the IPS by different amplitudes by intervalles
 in inverse scale ($k\;\rm in\; h Mpc^{-1}$). Following the different works,
 one finds that the best shape (as needed to fit the WMAP CMB angular
 power spectrum) is in agreement with a power law (Briddle et
 al. 2003) or present a bent at a particular scale (see Mukherjee et
 al. 2003 and figure 1) around $k \sim 0.01 \rm  Mpc^{-1}$.

\begin{figure}[h]
   \centering
     \includegraphics[width=4.5cm]{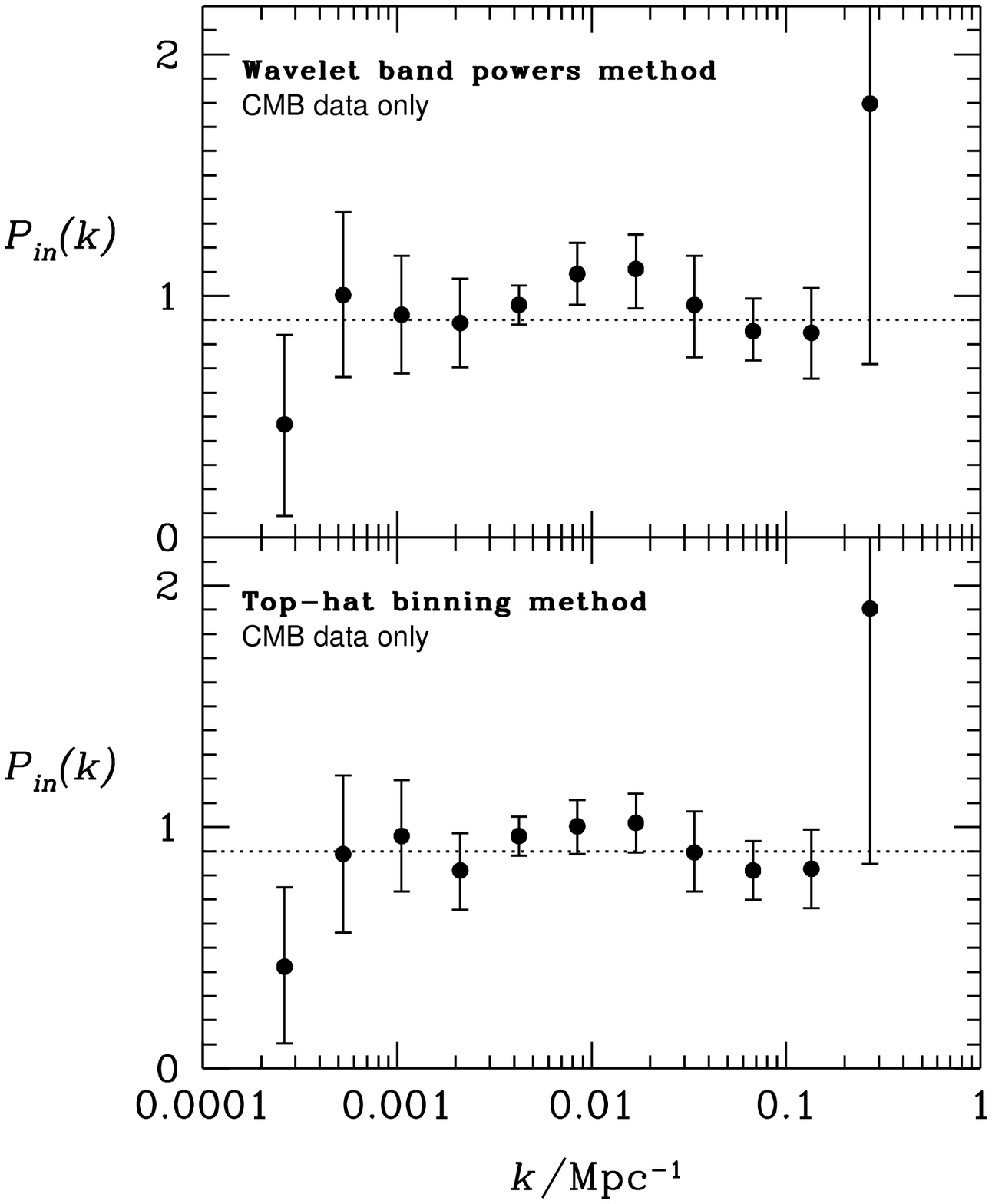}\includegraphics[width=4.5cm]{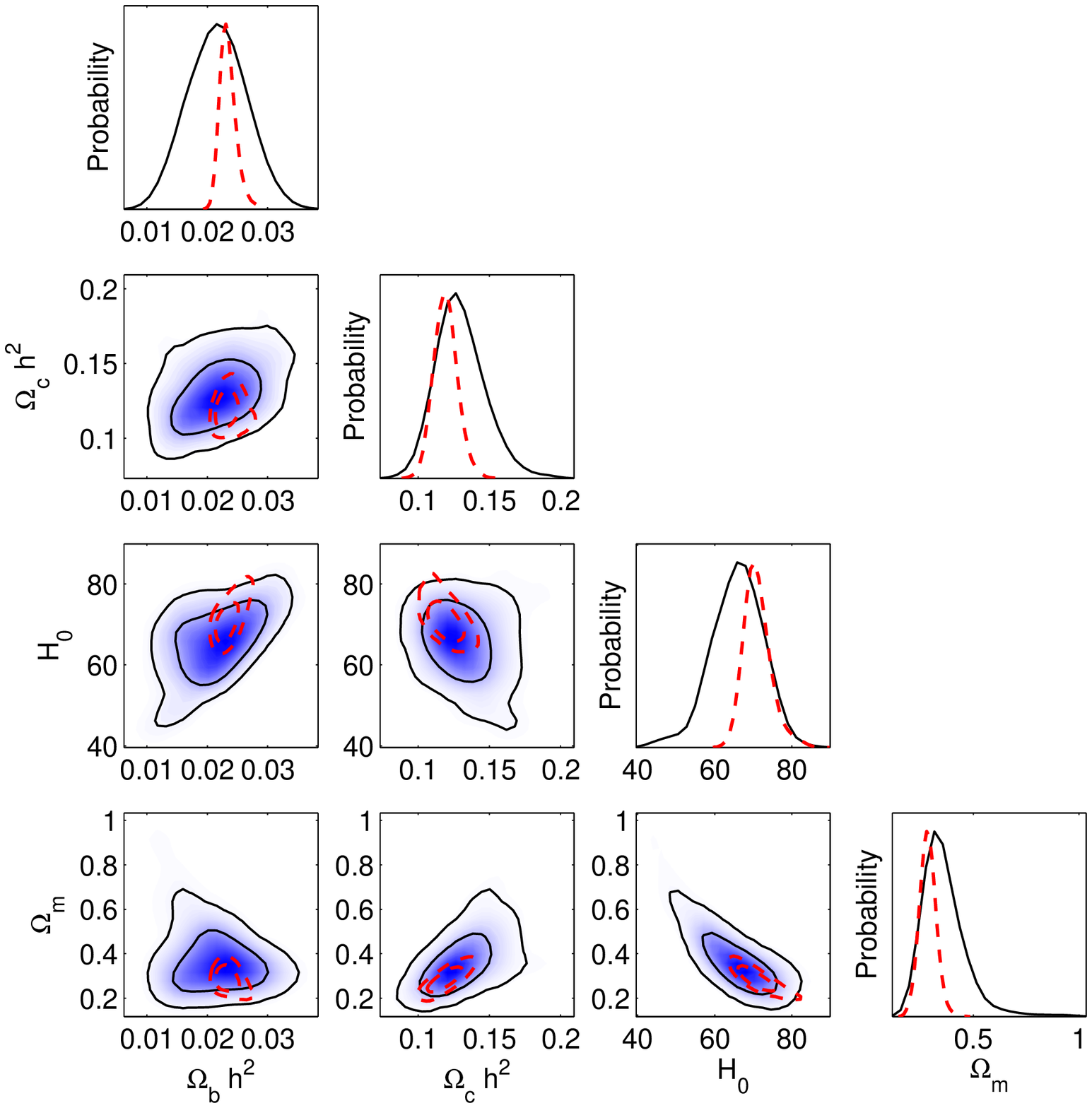}
      \caption{Left: shape of the IPS when parametrized in amplitudes in bins for WMAP fitting (from Mukherjee et al. 2003). 
       Right: Effect of such parametrization on the CP constraints: in red the likelihoods with power law IPS and in blue assuming free amplitudes in bins (from Briddle et al. 2003) }
       \label{figure_mafig}
   \end{figure}

 \subsection{Broken initial spectrum}

A second parametrization considered is the broken initial power
spectrum. The idea is to split the spectrum in two parts, two power
laws having independent indexes caracterising large and small
scales. Blanchard et al. (2003) have shown that in this situation WMAP
angular power spectrum (in temperature and cross temperature
polarization cases), large scale surveys (2dF, APM, PSCz) and Lyman
alpha inferred power spectra are fully in agreement with an Einstein de Sitter
model. Figure 2 shows that even Planck will be unable to make the
difference between a $\Lambda$-CDM power law cosmology and a
$\Omega_{matter}=1$ model with a broken power spectrum.

\begin{figure}[h]
   \centering
\includegraphics[width=4.5cm]{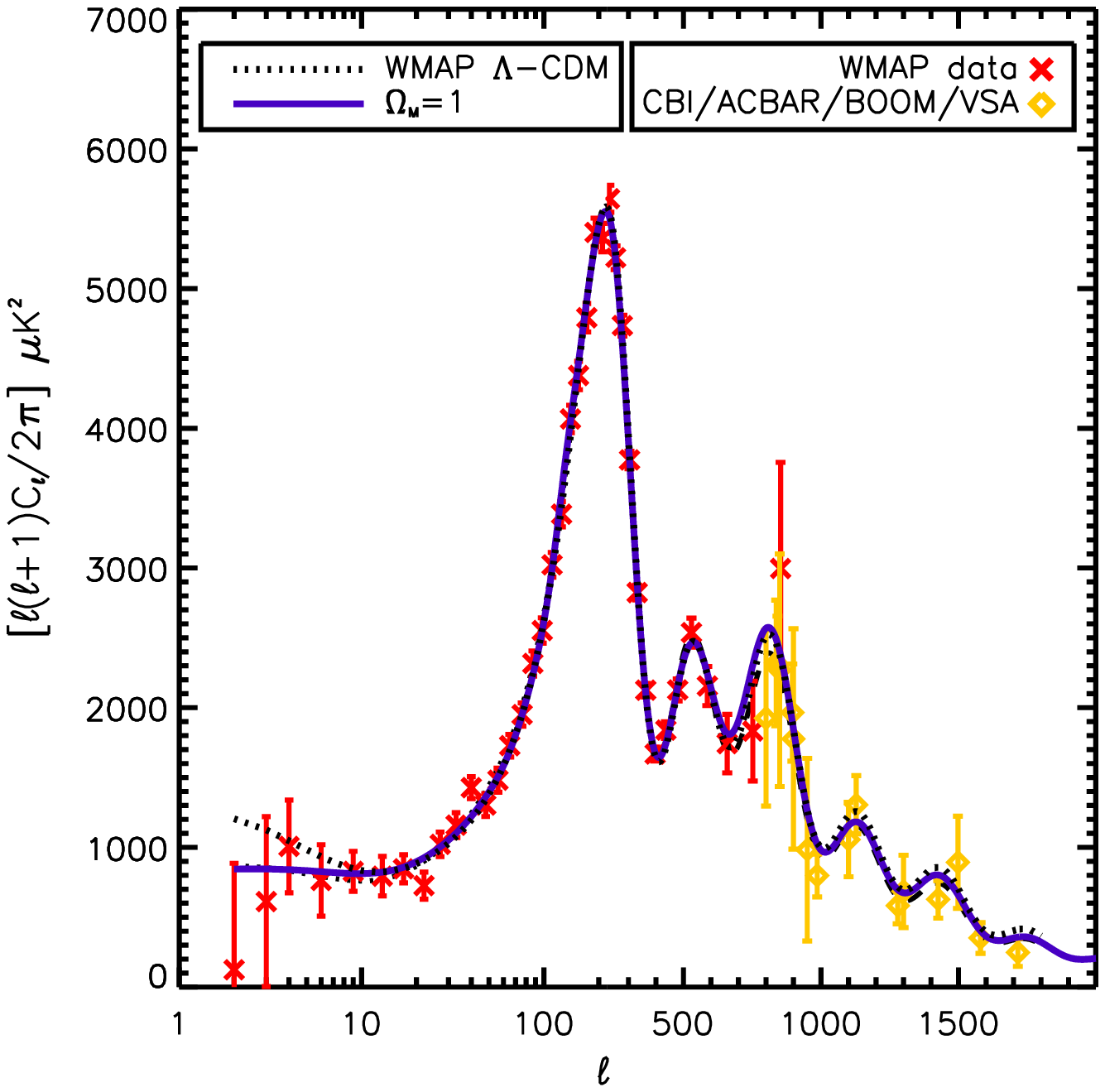}   \includegraphics[width=4.5cm, height=4.5cm]{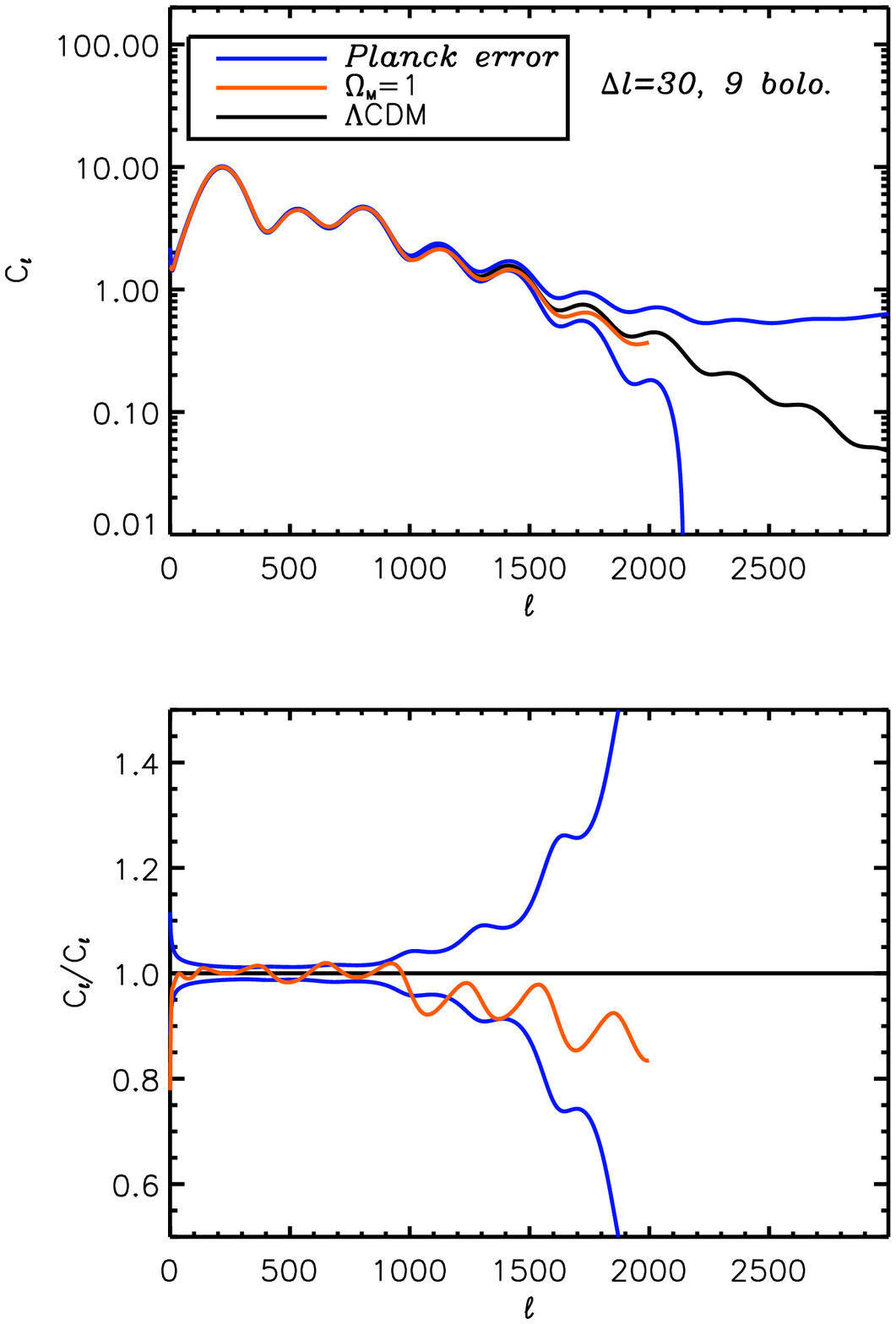} 
	\caption{$\Lambda-CDM$
   power law angular spectrum and $\Omega_{matter}=1$ model
   over-plotted on WMAP data and their difference compared to Planck
   estimated error bars}
 \label{figure_mafig2} 
\end{figure}

 \subsection{Reconstructing the shape}
 
Having such good datasets as WMAP allows one to probe the initial
condition of cosmic scenarios. The angular power spectrum of CMB
anisotropies is a convolution of ``late time cosmology'' effects
(transfer function, $\Delta$) with initial conditions (IPS of
fluctuations, $P^0$ ): $C_{\ell}=4\pi\int\Delta^{2}_{\ell}(k)\, P^0(k)
\, \frac{\rm{d} k}{k} \simeq WP^{0}$, where the last term is in matrix
notation and represents a numerical approximation to the integral.
Generally speaking, it is not possible to invert the matrix $W$ to
solve for the power spectrum, because each $C_{\ell}$ embraces
information about a limited range in $k$-space. However the inversion
is feasible under some assumptions. Basically the lack of information
can be remedied by the introduction of priors, such as for example the
requirement of a certain degree of smoothness in the
solution. Tocchini Valentini, Douspis and Silk (2004) have presented a
new method reaching this goal, with the advantage of knowing the error
bars on the reconstructed IPS. Other methods have been proposed in the
same spirit (Shafieloo et al. 2004; Kogo et al. 2004).

Figure 3 (left) shows the reconstructed power spectrum from WMAP
individual $C_{\ell}$'s when ``concordance'' late time cosmological
parameters are assumed. The horizontal black line represent the best
fitting power law IPS found in Spergel et al. 2003. The first remark
is perhaps that the reconstructed IPS is not far from a power
law. Nevertheless, deviations occur and are needed in order to fit
features seen in the angular power spectrum and responsible for the
``bad'' goodness of fit of a power law spectrum. The reconstructed IPS
allows one to fit the WMAP $C_{\ell}$'s in improving by 44 the
log-likelihood ($\sim \chi^2$) compared to the power law
$\Lambda$-CDM. Such spectrum let some imprints in the angular power
spectrum but also in the matter power spectrum. Recent surveys probing
large scales are then able to test such deviations. Figure 3 (right)
shows the comparison between the IPS found in Fig.~3 (left) evolved
to now, and the matter power spectrum observed by the Sloan Digital
Sky Survey (SDSS). More than being compatible, here again, the
reconstructed IPS allows to improve the goodness of fit, by fitting
nicely the deviations seen in SDSS data. 
\begin{figure}[h]
   \centering
   \hspace{-4.3cm}\includegraphics[width=4cm, height=2.3cm]{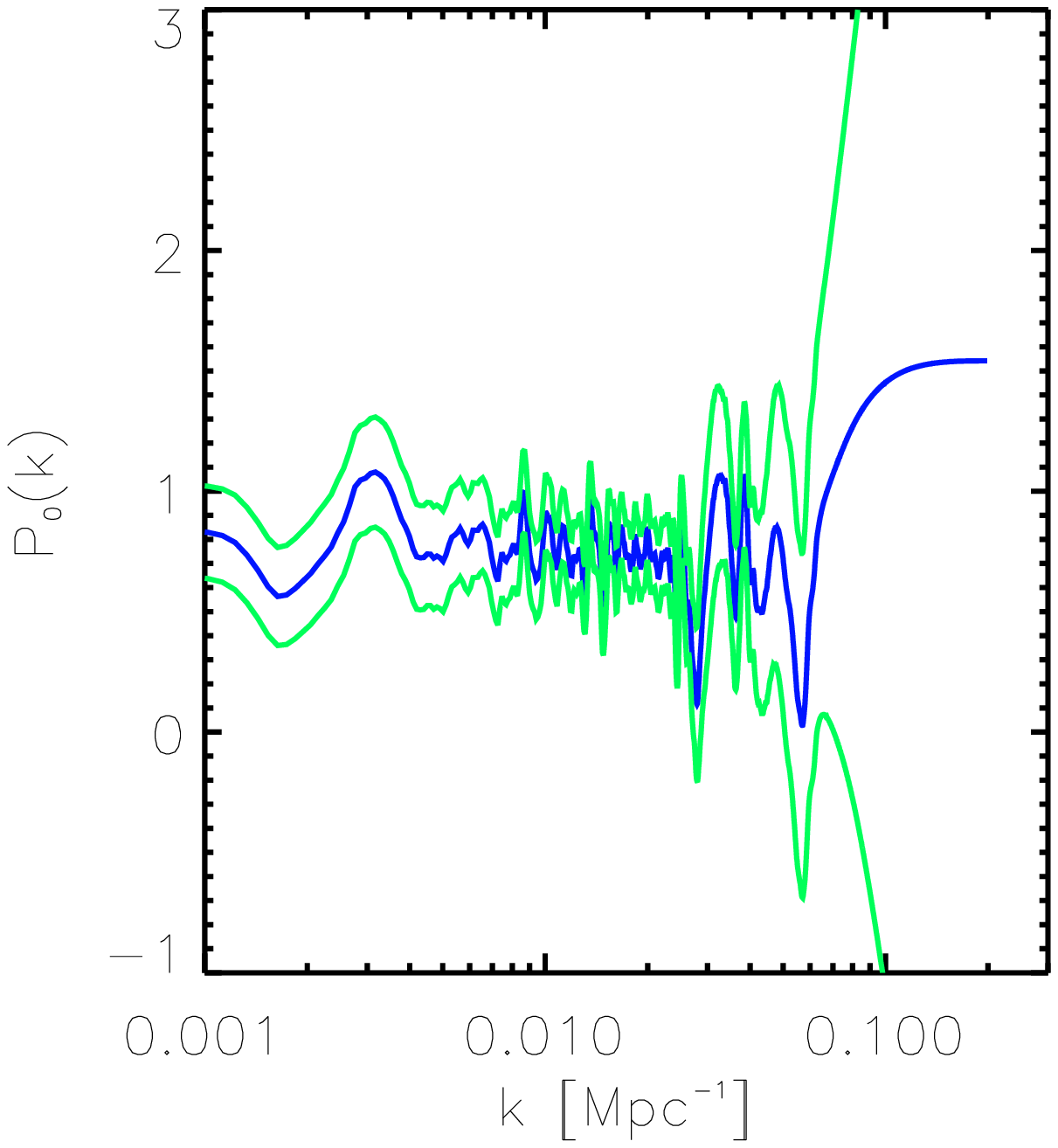}
   \hspace{2.5cm}\includegraphics[width=6cm,
   height= 2.3cm]{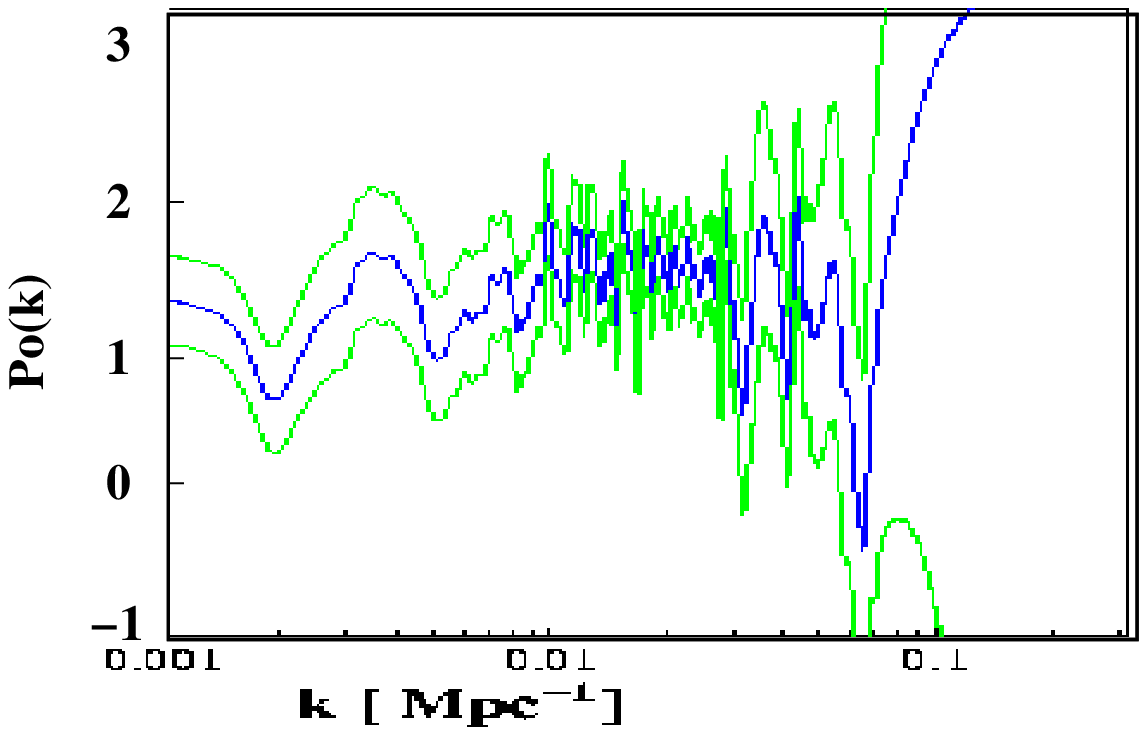}\hspace{-8cm}\includegraphics[width=3.5cm]{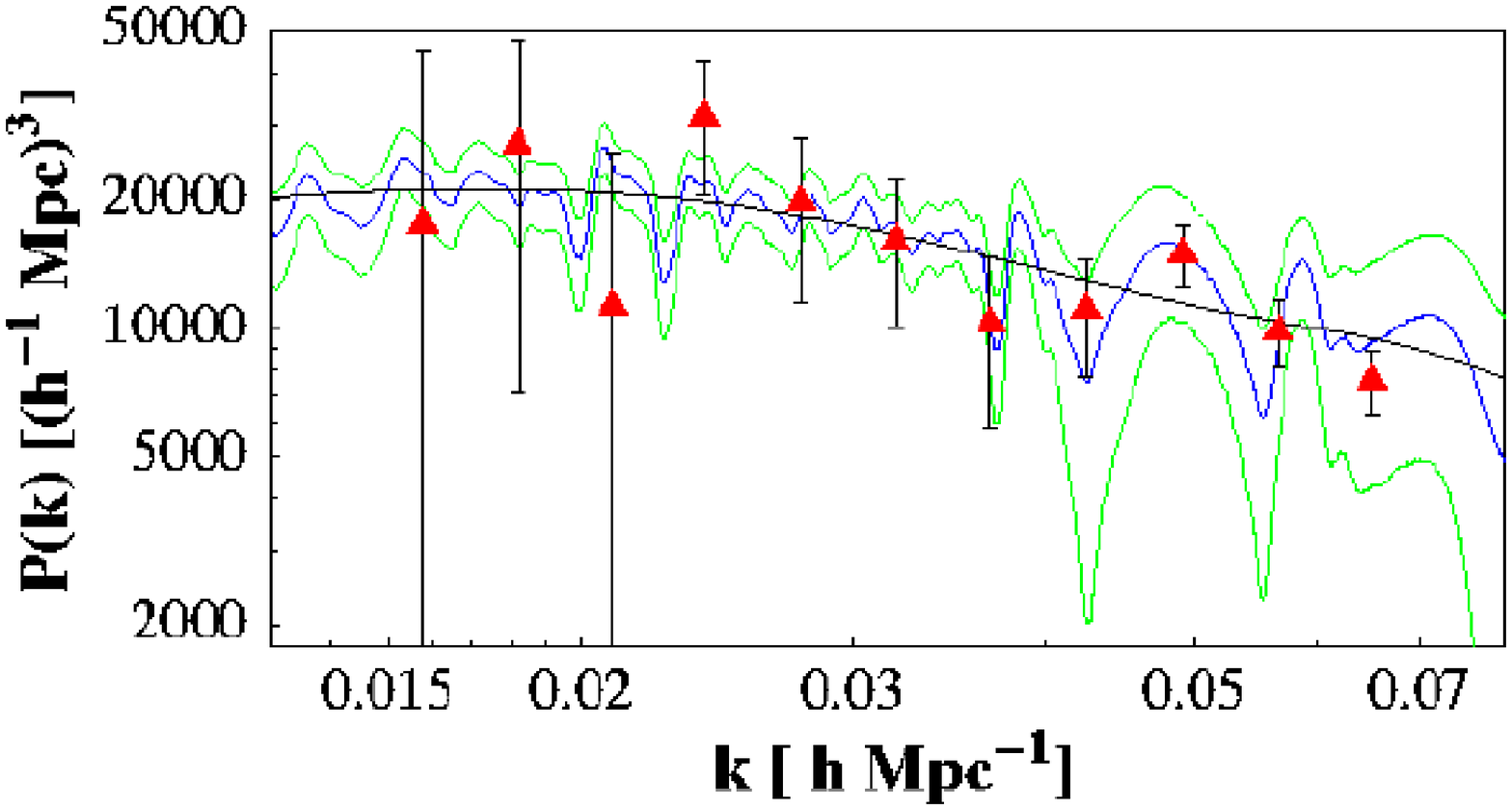}
	 \caption{Reconstructed IPS from
   WMAP under concordance prior (left) -- with corresponding matter power spectrum compared to SDSS
   (center)-- and Einstein de Sitter prior
   (right); from Tocchini-Valentini et al. 2004. } \label{figure_mafig} 
\end{figure} 
If such deviations (in the angular and matter PS) remain in the future
(at this level), it will be hard firstly to detect them in the IPS (if
they are not systematics effects) and then to explain such behavior
(especially if they are not coherent; see Martin \& Ringeval (2004) for
coherent oscillations in the IPS). Furthermore, if the ``concordance''
cosmological parameter prior is not set, one is able to find the
``best'' IPS fitting the data for almost any kind of cosmological
parameters given the strong degeneracy between initial conditions and
late time cosmology. Figure 6 shows for example the IPS reconstructed
from WMAP $C_{\ell}$'s by forcing $\Omega_{Matter}=1$ and the
cosmological parameters presented in Blanchard et al. (2003). The
corresponding angular spectrum gives by definition an acceptable
goodness of fit ($\chi^2/(degrees\; of\; freedom) \sim 1$).


\section{Conclusion}

The degeneracy between primordial cosmology, IPS,
and late time cosmology, CP,  allows one to mimic
actual (and future) CMB angular power spectrum observations. If the
hypothesis of unique power law initial power spectrum is relaxed, the
constraints from CMB (and large scale surveys) on cosmological
parameters are weakened. Such degeneracy shows then the need for other
probes and combinations (clusters of galaxy, supernovae, lensed
galaxies, ...) in order to reach precision cosmology and to set strong
constraints on cosmic scenarios.



\begin{thebibliography}{}

\bibitem{} Blanchard, A., Douspis, M., Rowan-Robinson, M., \& Sarkar, S.\ 2003, A\&A, 412, 35 

\bibitem{} Bridle, S.~L., Lewis, A.~M., Weller, J., \& Efstathiou, G.~P.\ 2003, New Astronomy Review, 47, 787 

\bibitem{} Kogo N. et al., 2004, ApJ, 607, 32

\bibitem{} Martin, J., \& Ringeval, C., 2004, PRD 69, 083515

\bibitem{} Mukherjee, P.~\& Wang, Y.\ 2003, ApJ, 599, 1 

\bibitem{} Shafieloo A. \& Souradeep T, 2004, PRD, 70, 043523

\bibitem{} Spergel, D.~N., et al.\ 2003, ApJs, 148, 175 

\bibitem{} Tocchini Valentini, D., Douspis, M. and Silk, J., MNRAS in press, astro-ph/0402583 



\end{thebibliography}
\end{document}